\documentclass[useAMS,usenatbib]{mn2e}

\usepackage{graphicx}

\title[The Accuracy of Subhalo Detection]{The Accuracy of Subhalo Detection}
\author[Muldrew, Pearce \& Power]{Stuart I. Muldrew$^{1}$\thanks{E-mail:
ppxsm2@nottingham.ac.uk}, Frazer R. Pearce$^{1}$ and Chris Power$^{2}$\\
$^{1}$School of Physics and Astronomy, University of Nottingham, Nottingham, NG7 2RD, UK\\
$^{2}$Department of Physics and Astronomy, University of Leicester, Leicester, LE1 7RH, UK}
\begin{document}

\date{Accepted 2010 September 3. Received 2010 August 31; in original form 2010 July 20}

\pagerange{\pageref{firstpage}--\pageref{lastpage}} \pubyear{2010}

\maketitle

\label{firstpage}

\begin{abstract}
With the ever increasing resolution of $N$-body simulations, accurate subhalo detection is becoming essential in the study of the formation of structure, the production of merger trees and the seeding of semi-analytic models.  To investigate the state of halo finders, we compare two different approaches to detecting subhaloes; the first based on overdensities in a halo and the second being adaptive mesh refinement.  A set of stable mock NFW dark matter haloes were produced and a subhalo was placed at different radii within a larger halo.  \textsc{subfind} (a Friends-of-Friends based finder) and \textsc{ahf} (an adaptive mesh based finder) were employed to recover the subhalo.  As expected, we found that the mass of the subhalo recovered by \textsc{subfind} has a strong dependence on the radial position and that neither halo finder can accurately recover the subhalo when it is very near the centre of the halo.  This radial dependence is shown to be related to the subhalo being truncated by the background density of the halo and originates due to the subhalo being defined as an overdensity.  If the subhalo size is instead determined using the peak of the circular velocity profile, a much more stable value is recovered.  The downside to this is that the maximum circular velocity is a poor measure of stripping and is affected by resolution.  For future halo finders to recover all the particles in a subhalo, a search of phase space will need to be introduced.
\end{abstract}


\section{Introduction}
\label{intro}

It has long been understood that dark matter plays an essential role in galaxy formation.  \citet{White78} demonstrated that dark matter haloes act as potential wells within which infalling material can be captured and condense to form galaxies.  As the universe ages, these haloes merge to form larger structures and this continued process produces the framework of the universe that we see today.  This so called hierarchical model of galaxy formation has been put to many tests including those generated by $N$-body simulation. One of the most widely used of these simulations is the \emph{Millennium Simulation} \citep{Springel05b} which accurately reproduced the large-scale structure of a 500 Mpc/$h$ cube region of the universe.

One of the challenges of studying the results of $N$-body simulations has been finding a consistent way of identifying the structures and substructures within them.  Detailed studies of haloes and subhaloes require halo finders, codes that scan the simulation outputs and identify structures.  Many different halo finders are available and each uses different techniques and definitions of the haloes they find.  Broadly, halo finders fall into two general categories; those based on the Friends-of-Friends (FoF) technique and those based on grids.

FoF was first proposed by \citet{Davis85} and locates haloes based on a predetermined linking length for particles.  This is usually a fraction of the mean inter-particle separation and any two particles closer than this distance are linked together.  Isolated sets of linked particles are then identified as the haloes.  Commonly a value of 0.2 times the mean inter-particle separation is chosen motivated by SCDM ($\Omega_{0}=1.0$ \& $\Omega_{\Lambda}=0.0$) \citep{Davis85} and a slightly lower value of 0.16 is sometimes adopted for $\Lambda$CDM ($\Omega_{0}=0.3$ \& $\Omega_{\Lambda}=0.7$) \citep{Lacey93, Eke96}.  Despite the difference, convergence between cosmologies in the halo mass function can be found using 0.2 \citep[see][]{Jenkins01} making this the most widely used.  The FoF method was implemented in, for example, \textsc{subfind} \citep{Springel01} and \textsc{hfof} \citep{Klypin99}, with different techniques being used to find subhaloes.  \textsc{hfof} uses hierarchical FoF to locate the subhaloes by using a shorter linking length inside the halo, while \textsc{subfind} searches the haloes for overdensities in the density profile.

The grids method of halo finding works by placing a grid across the simulation and smoothing the discrete particle data onto that grid and then locating the densest cells.  Refinement can be built onto the grid to obtain improved resolution and to increase the speed of the code.  The density peaks that are located on the grid can then be used as the seeds for potential structures.  This technique was used by, for example, \textsc{ahf} \citep{Knollmann09} and \textsc{asohf} \citep{Planelles10}.  The variations between these codes comes in the definition of haloes.  \textsc{ahf} uses isodensity contours on the grid, while \textsc{asohf} uses spherical overdensities.

FoF and grid based methods are the two main ways for locating structure, but there are alternatives.  More recent finders, such as \textsc{hsf} \citep{Maciejewski09}, have tried using phase space to identify subhaloes.  This extends the search based on position and density to incorporate the velocity of the particles.  Bulk velocities can then also be used to help identify structures.  Other finders that have tried different techniques include \textsc{voboz} \citep{Neyrinck05}, which replaced the uniform grid with a Voronoi diagram, and \textsc{surv} \citep{Tormen04, Giocoli10}, which uses knowledge of the structures from one snapshot to help find structure in the next.  While this summary of halo finders is by no means exhaustive, it does give a flavour for the different techniques employed.  A thorough review of the different types of halo finders available and their effectiveness will be found in Knebe et al. (in preparation).

The importance of accurate subhalo detection has increased in recent years with the advances in high resolution simulations.  Various simulations of Milky Way sized haloes have been produced including \emph{via Lactea} \citep{Diemand07, Diemand08}, \emph{Aquarius} \citep{Springel08} and \textsc{ghalo} \citep{Stadel09}.  As expected, these haloes contain a wealth of substructure \citep[see][]{Gao04}.  However, it is important to ask how robust the recovered properties of subhaloes are to the choice of subhalo finder.  For example, subhaloes are identified initially as overdensities in their host haloes.  We expect picking out such overdensities to be more difficult in the innermost parts of the host haloes where the background density is the greatest.  If one halo finder is less able to pick out these overdensities than another halo finder, we would expect this halo finder to systematically underpredict the numbers of subhaloes in the inner parts of haloes, which would have important implications for how we interpret the results of, for example, the radial distribution of subhaloes and subhalo mass loss.

In this paper we set out to quantify the extent to which our choice of halo finder impacts on the radial distribution of subhaloes that we recover.  Specifically we focus on \textsc{subfind} \citep{Springel01} and \textsc{ahf} \citep{Knollmann09} and ask how well these halo finders can recover the properties of a NFW subhalo \citep{Navarro97} embedded in a more massive host NFW halo.  The advantage of this approach is that, unlike using haloes and subhaloes drawn from cosmological simulations, we know exactly which particles belong to the host and to the subhalo at initial time and we can track their positions and velocities at all subsequent times.  This provides a clean test of the halo finders because any discrepancies found can be identified easily.

The rest of this paper is setout as follows. In \S\ref{method} we outline the methods used, including summaries of the halo finders and the process of constructing a mock 6D ($x$, $y$, $z$, $v_{x}$, $v_{y}$, $v_{z}$) NFW halo by reproducing the density and velocity profiles.  We then use this construction, in \S\ref{model}, to model an infalling subhalo.  This is undertaken in two ways, first by considering how well the halo finders recover the subhalo when simply placed at different radii within the main halo.  The second method is to let the subhalo fall into the main halo under gravity and compare how the different halo finders recover the subhalo.  Having established the accuracy of the halo finders, in \S\ref{strip} we investigate the effect the trajectory of the subhalo has on stripping as it passes through the halo.  In \S\ref{circ} we test the reliability of recovering the peak in the circular velocity profile.  Finally we summarise our results.  Throughout this work, a standard $\Lambda$CDM cosmology has been adopted, taking $\Omega_{0}=0.3$, $\Omega_{\Lambda}=0.7$ and $h=0.73$, where appropriate, consistent with observations from \textsc{wmap} first year results \citep{Spergel03}.

\section{Methods}
\label{method}

\subsection{Halo Finders}
\label{halofin}

For the purpose of this work we focus on two halo finders that rely on different methods to detect haloes and subhaloes.

\subsubsection{\textsc{ahf}}
\label{ahf}

\textsc{ahf}\footnotemark \citep{Knollmann09} is an updated version of \textsc{mhf} \citep{Gill04} and works using an adaptive mesh refinement method.  It begins by placing a user-defined grid across the box and calculates the particle density in each cell.  If this is greater than a user-specified value, then the cell is refined with a smaller grid.  The particle density is then recalculated on this finer grid and, if required, further refinement is carried out.  Once all the refinements are carried out, a hierarchical grid tree of the density distribution has been produced and this can be used to find structure.  Throughout this work, we used a grid of 128 cells with refinement being carried out in cells that contain more than 3 particles.

\footnotetext{Available from http://popia.ft.uam.es/AMIGA}

The most refined and isolated cells are used as potential halo centres and these are linked to the coarser grids to build the structure.  If two isolated centres join up on a coarser grid then these are combined into one structure.  By considering these separate, isolated points in one structure, substructure can be defined.  Once the structures are identified, starting on the lowest level of substructure, they are tested for boundness in isolation.  This is conducted by comparing the particles velocity to the local escape velocity obtained using a spherical potential approximation.  If a particle is found to be unbound it is assigned to the next highest level of structure until it is dispensed with if not bound to the halo.  The haloes are then truncated at the virial radius (see \S\ref{mock}) to define their size.  For the subhaloes, not all have a low enough overdensity to satisfy the virial radius due to the background density of the halo.  If this is the case then they are truncated by a sharp spherical boundary at the outer radius at which their density profile first shows an upturn and starts to rise with increasing distance.

\subsubsection{\textsc{subfind}}
\label{sub}

\textsc{subfind} \citep{Springel01} begins by conducting a standard Friends-of-Friends (FoF) search of the simulation volume to identify haloes.  At each particle the local density is then calculated using a local SPH-like smoothing kernel interpolation over the nearest neighbours.  Any locally overdense region is then considered as a subhalo candidate with its shape being defined by an isodensity contour that traverses the saddle point in the density profile of the halo.  This is found by lowering the global density threshold and selecting out the overdense regions.  At this stage particles can be members of more than one structure allowing different levels of substructure to be determined.  For this work, we used a FoF linking length of 0.2 and 10 particles for the SPH density calculation allowing \textsc{subfind} to recover all subhaloes with 10 or more particles.  Tests were also carried out using higher values for the SPH density calculation, but the number of particles recovered was found to be relatively insensitive to this parameter for the size of the subhalo we used.

Once subhalo candidates have been identified, an unbinding procedure is used to determine iteratively which particles are not gravitationally bound.  This is achieved by defining the centre of the subhalo as the position of the most bound particle and the bulk velocity as the mean velocity of the particles in the group.  The kinetic and potential energies of the particles are then compared and unbound particles are removed.  The final step is to assign particles that are listed in multiple structures to just one.  To solve this, the particles are assigned to the smallest structure they are found in.  The remaining FoF particles that have not been assigned to substructure are then tested for boundness and assigned to the background halo.  Any particles that are not bound to anything are then classified as FoF 'fuzz'.

\subsection{Constructing a Mock Halo}
\label{mock}

The following outlines the process of constructing a mock dark matter halo.  For simplicity we have limited ourselves to the case of a spherical halo that follows a radial NFW density profile,
\begin{equation}
 \label{NFWdens}
 \rho(r) =\frac{\rho_{\rm crit}\delta_{\rm c}}{r/r_{\rm s}(1+r/r_{\rm s})^2},
\end{equation}
\noindent where $\rho_{\rm crit}$ is the critical density of the universe, $r_{\rm s}$ is the scale radius and $\delta_{\rm c}$ is the characteristic density.  Dark matter haloes are characterised by their virial mass,
\begin{equation}
 M_{\rm vir}=\frac{4\pi}{3}r^3_{\rm vir}\Delta_{\rm vir}\rho_{\rm crit},
\end{equation}
\noindent where $r_{\rm vir}$ is the virial radius and $\Delta_{\rm vir}$ is the virial approximation given by \citet{Bryan98} as,
\begin{equation}
 \label{vir}
 \Delta_{\rm vir} = 18\pi^2+82(\Omega(z)-1)-39(\Omega(z)-1)^2,
\end{equation}
\noindent where,
\begin{equation}
 \label{omegaz}
 \Omega(z) = \frac{\Omega_{0}(1+z)^3}{\Omega_{0}(1+z)^3+\Omega_{\Lambda}}.
\end{equation}
\noindent For $\Omega_{0}=0.3$, $\Omega_{\Lambda}=0.7$ and $z=0.0$, $\Delta_{\rm vir}\approx101$.  Using the scale radius and the virial approximation, the characteristic density is given by,
\begin{equation}
 \label{deltac}
 \delta_{\rm c} =\frac{\Delta_{\rm vir}}{3}\frac{c^3}{\ln(1+c)-c/(1+c)},
\end{equation}
\noindent where $c = r_{\rm vir}/r_{\rm s}$ is the concentration.

Using these conditions, a Monte Carlo realisation can be constructed by defining the number of particles within $r_{\rm vir}$, $N_{\rm vir}$, and specifying the concentration of the halo required.  The Monte Carlo realisation is produced by drawing a random enclosed mass and inverting to find a radius.  This is then turned into a set of coordinates by specifying they produce a smooth distribution on the surface of a sphere.  The mass of a NFW halo continues to increase with increasing radius and so in principle has infinite mass; we circumvent this by truncating the halo beyond a cut-off radius, $r_{\rm cut}$.  This modifies the density profile so that $\rho(r<r_{\rm cut})$ follows the NFW profile and $\rho(r>r_{\rm cut})=0$.  For this work we set $r_{\rm cut}=2r_{\rm vir}$.  A smoother truncation could be produced by using a exponential decay at the edge of the halo. 

Once the halo is constructed, each particle needs to be given a velocity that reproduces the velocity dispersion, $\sigma(\rm r)$, of a halo.  Dark matter haloes are supported by the random motion of the particles and to get an accurate representation we need to reproduce this in the velocity of the particles.  The velocity dispersion can be obtained by considering the Jeans equation,
\begin{equation}
 \label{Jeans}
 \frac{1}{\rho}\frac{d}{dr}(\rho\sigma_{\rm r}^2)+2\beta\frac{\sigma_{\rm r}^2}{r} = -\frac{d\Phi}{dr},
\end{equation}
\noindent where $\beta=1-\sigma_{\theta}^2(r)/\sigma_{\rm r}^2(r)$ and $\Phi$ is the gravitational potential.  Assuming isotropy, $\sigma_{\theta}(r)=\sigma_{\rm r}(r)$, $\beta=0$ and the velocity dispersion is given by,
\begin{equation}
 \label{disperint}
 \sigma_{\rm r}^2(r)=\frac{1}{\rho(r)}\int^\infty_{r}\rho(r')\frac{d\Phi}{dr'}dr'.
\end{equation} 
\noindent This integral was solved by \citet{Lokas01}, and confirmed here, to give,
\begin{eqnarray}
 \label{disper}
 \frac{\sigma_{\rm r}^2}{V_{\rm vir}^2}
 &=& \frac{c^2s(1 + c s)^2}{2[\ln(1+c)-c/(1+c)]} \ [\pi^2-\ln(cs)-\frac{1}{cs} \nonumber \\
 &\hspace{-1.2cm} - & \hspace{-0.9cm} \frac{1}{(1+cs)^2} -\frac{6}{1+cs}
 + \left(1+\frac{1}{c^2s^2} - \frac{4}{cs}
 - \frac{2}{1+c s} \right) \nonumber  \\
 &\hspace{-1.2cm} \times & \hspace{-0.9cm}  \ln (1+c s)
 + 3 \ln^2 (1+c s) + 6 \, {\rm Li}_2(-cs) ] \ ,
\end{eqnarray}
\noindent where $s=r/r_{\rm vir}$, $V_{\rm vir}$ is the circular velocity at the virial radius and ${\rm Li}_{2}(x)$ is the dilogarithm function given by, \footnotemark

\footnotetext{Note that the dilogarithm approximation given in equation (17) of \citet{Lokas01} is not suitable for this task.}

\begin{equation}
 \label{dilog}
 {\rm Li}_{2}(x) = \int^0_{x}\frac{\ln(1-t)}{t}dt.
\end{equation}
\noindent The 3D velocity dispersion is then given by the sum of the individual components.  Since isotropy was assumed this gives $\sigma^2_{\rm 3D}(r)=3\sigma^2_{\rm r}(r)$.  To generate a velocity distribution function for a given radius, a Maxwell-Boltzmann distribution can be assumed \citep[cf.][]{Hernquist93},
\begin{equation}
 \label{MB}
 F(v,r) = 4\pi\left(\frac{1}{2\pi\sigma^2_{r}}\right)^{3/2}v^2\exp\left(\frac{-v^2}{2\sigma^2_{r}}\right).
\end{equation}
\noindent The function $F(v,r)$ is normalised such that,
\begin{equation}
 \label{norm}
 \int^\infty_{0}F(v,r)dv = 1.
\end{equation}
\noindent  The velocity of each particle can then be obtained using the probability distribution of equation~(\ref{MB}).  Having obtained the density and velocity profiles of the halo, the only thing left is to assign a direction to each velocity.  This is done by simply requiring that the directional velocity vectors produce a smooth distribution on the surface of a unit sphere. 

\begin{figure}
\includegraphics[width=82mm]{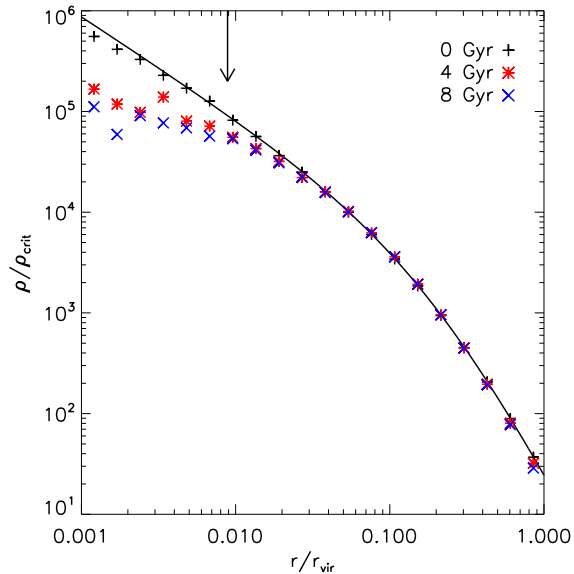}
\caption{The density profile of a $M_{\rm vir}=10^{14} \rm M_{\odot}$, $N_{\rm vir}=10^6$ and $c=5$ halo left to evolve over 8 Gyr.  The black line denotes the theoretical NFW profile, while the mock halo is shown initially (black pluses), after 4 Gyr (red asterisks) and 8 Gyr (blue crosses).  The arrow represents the Plummer equivalent softening ($h=2.8\epsilon=8.4 \rm kpc$).}
\label{dens}
\end{figure}

To test the stability of this setup, an isolated halo with $M_{\rm vir}=10^{14} \rm M_{\odot}$, $N_{\rm vir}=10^6$ and $c=5$ was left to evolve over 8 Gyr using \textsc{gadget}-2 \citep{Springel05}.  The spline gravitational softening was set to $\epsilon=3 \rm kpc$ corresponding roughly to the radius of the 100th particle \citep[see][]{Power03}.  Fig.~\ref{dens} shows that the halo retains the overall shape of an NFW profile, except at the centre where the profile has flattened similar to that observed by \citet*{Kazantzidis04}.  This flattening of the density profile is caused by approximating the distribution function with a Maxwell-Boltzmann.  As demonstrated in \citet{Kazantzidis04}, this will lead to an over estimate of any stripping that occurs.  Despite this, it will have no effect on the ability of halo finders to recover the haloes.  This was confirmed by using the method outlined in \citet{Read06} to generate haloes with \citet{Plummer11} and \citet{Hernquist90} density profiles based on their 6D distribution functions.  When the same tests were carried out on these haloes, the same patterns between the halo finders was found as for the NFW with the Maxwell-Boltzmann approximation.

\section{Modelling an Infalling Subhalo}
\label{model}

\subsection{Static Infall}
\label{place}

The first method of modelling the infall of a subhalo we adopted was to consider how well different halo finders recovered the subhalo at a given radius.  This was achieved by placing the same sized subhalo \emph{by hand} at different radii within the main halo and attempting to recover it with each halo finder.  A halo was generated with $M_{\rm vir}=10^{14} \rm M_{\odot}$, $N_{\rm vir}=10^6$ and $c=5$ and a subhalo with $M_{\rm vir}=10^{12} \rm M_{\odot}$, $N_{\rm vir}=10^4$ and $c=12$.  The concentration of the subhalo was set to be higher than the halo in order to reflect the conditions found in cosmological simulations \citep[see][]{Bullock01, Eke01}.  The subhalo was then placed at different distances away from the centre of the halo and given a velocity,
\begin{equation}
\label{gp}
v = \sqrt{\frac{2GM_{\rm halo}}{r_{\rm sep}}},
\end{equation}
\noindent where $M_{\rm halo}$ is the mass of the halo and $r_{\rm sep}$ is the separation of the centres of the halo and subhalo, towards the centre of the halo.  This velocity corresponds to the conversion of potential energy to kinetic, for two point masses, as the subhalo falls in from infinity.  When the subhalo was placed at the centre of the halo, $r_{\rm sep}=0.0$ so $v\rightarrow\infty$.  To overcome this, the subhalo was given a velocity of the previous closest separation when it was at the centre of the halo.  This set-up was produced 100 times for each separation using different random number seeds.  Consistent realisations were found each time.

\begin{figure}
\includegraphics[width=82mm]{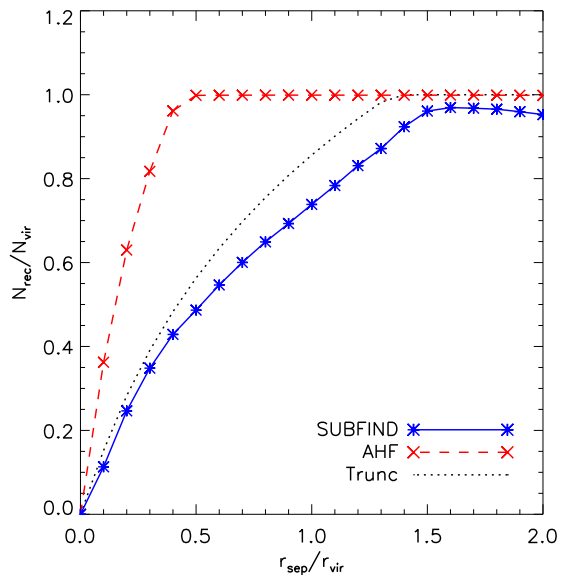}
\caption{The fraction of particles recovered at a given separation as the subhalo is placed at different positions within the halo.  Both halo finders recover consistent sizes across the multiple realisations, resulting in small error bars.  The dotted line represents the fraction of particles recovered if the subhalo is truncated at the radius where its density is equal to the background density of the halo.}
\label{subpos}
\end{figure}

Fig.~\ref{subpos} shows the fraction of particles recovered by each halo finder at different separations.  Neither halo finder can recover the subhalo when it is near the centre of the halo.  This corresponds to the densest region of the halo and leads to any overdensity from the subhalo being hidden.  As the separation is increased \textsc{ahf} has a steep rise in the fraction of particles it recovers until it is finding the complete subhalo from $\sim0.5r_{\rm vir}$ outwards.  \textsc{subfind} does not have such a drastic change and continues to underestimate the size of subhalo all the way out to $\sim1.5r_{\rm vir}$.

\begin{figure}
\includegraphics[width=82mm]{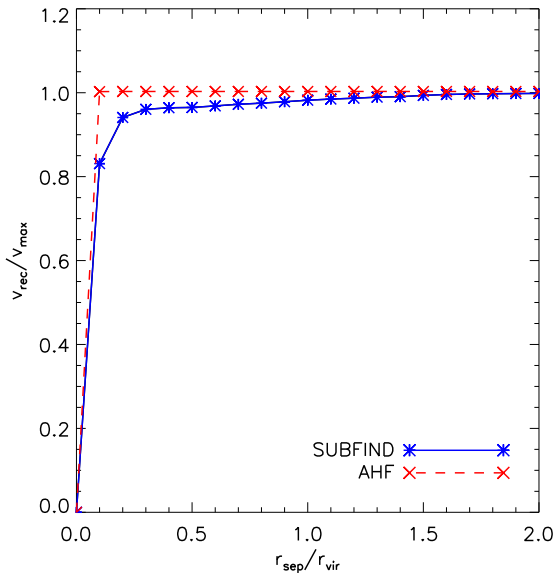}
\caption{The maximum circular velocity of the recovered subhalo as it is placed at different separations.  Both halo finders accurately recover the peak, with a small radial dependence displayed in \textsc{subfind}.}
\label{vc}
\end{figure}

We can gain some insight into the strong radial dependence in recovered particle number in \textsc{subfind} by considering the following simple argument.  \textsc{subfind} identifies subhaloes as overdensities; it identifies when a subhalo's local density equals its host halo's local density.  This equates to,
\begin{equation}
\label{eqnfw}
\frac{\delta_{c_{\rm sub}}}{\frac{r}{r_{s_{\rm sub}}}\left(1+\frac{r}{r_{s_{\rm sub}}}\right)^2}=\frac{\delta_{c_{\rm halo}}}{\frac{r_{\rm sep}-r}{r_{s_{\rm halo}}}\left(1+\frac{r_{\rm sep}-r}{r_{s_{\rm halo}}}\right)^2},
\end{equation}
\noindent where $\delta_{c_{\rm halo}}$ and $\delta_{c_{\rm sub}}$ are the characteristic densities of the halo and subhalo respectively (equation~\ref{deltac}), $r_{s_{\rm halo}}$ and $r_{s_{\rm sub}}$ are the scale radii of the halo and subhalo respectively, $r_{\rm sep}$ is the separation of the centres of the halo and subhalo and $r$ is the radius of the subhalo at which the densities are equal.  The number of particles within $r$ cannot exceed $N_{\rm vir}$ by construction.  The shape of the theoretical curve (dotted line in Fig.~\ref{subpos}) implied by equation~(\ref{eqnfw}) reasonably captures the shape of the curve recovered by \textsc{subfind}.  The agreement is not perfect, equation~(\ref{eqnfw}) predicts more mass should be recovered at larger radii than is recovered in practise, but the differences can be easily understood.  First, based on the random nature of the velocity assignment some of the particles will have large velocities and will therefore not be bound.  The effect of this will be to cause the two curves to deviate systematically from each other with increasing radius.  Second, \textsc{subfind} identifies overdensities as saddle points in the mass density profile rather than by equating subhalo and halo mass profiles, as implied by equation~(\ref{eqnfw}).  Overall the curve shares the same shape as that found using \textsc{subfind}, indicating that the background density is affecting the ability to recover the subhalo.

Implanting a NFW subhalo in a larger halo, defining the virial radius using equation~(\ref{vir}), is obviously a highly idealised situation.  Realistically the subhalo would be expected to undergo stripping which would cause it to be stripped down to its tidal radius at different points within the halo.  This tidal radius would roughly correspond to the radius at which there is a saddle point in the density profile \citep*{Tormen98}.  This also corresponds to the size of the overdensity that \textsc{subfind} is recovering.  Therefore, if the edge of the subhalo is defined as the tidal radius, \textsc{subfind} would give consistent recovery of the subhalo. 

\begin{figure}
\includegraphics[width=82mm]{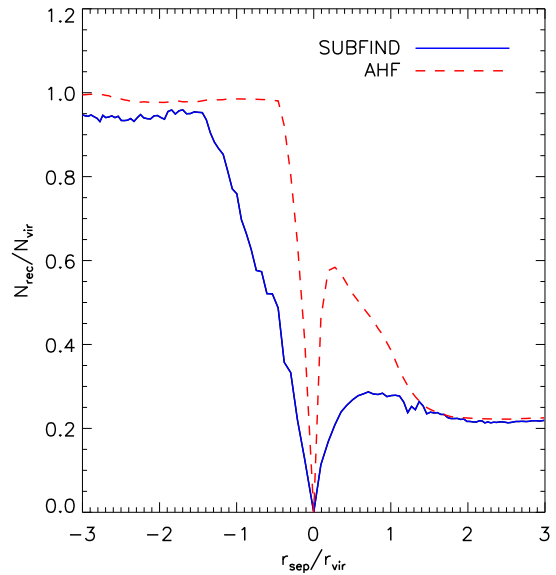}
\caption{The fraction of particles recovered at a given radius as the subhalo is allowed to fall into a halo from infinity.  The subhalo experiences the most stripping when it passes through the centre of the halo.  Neither halo finder can detect the subhalo as it passes through the centre of the halo and they yield different sizes for the subhalo either side of this region.}
\label{infall}
\end{figure}

A different method of determining the size of the subhalo is to consider the peak in the circular velocity profile \citep[see][]{Ghigna98, Ghigna00}.  This will be less affected by truncation of the subhalo, as the particle with the maximum circular velocity is closer to the centre.  Fig.~\ref{vc} shows the recovered maximum circular velocity for the subhalo at different separations.  This was obtained by calculating the circular velocity for each particle in the subhalo and taking the largest of these as the peak.  As expected, both halo finders more accurately recover the subhalo size using this method.  \textsc{subfind} still displays a slight radial dependence, with a gradual decrease towards the centre of the halo.  This is caused by high velocity particles near the centre of the subhalo being unbound due to the truncation.  As the subhalo was not detected at the centre of the halo, it is not possible to obtain a circular velocity there. 

\subsection{Dynamic Infall}
\label{dynam}

\begin{figure*}
\includegraphics[width=150mm]{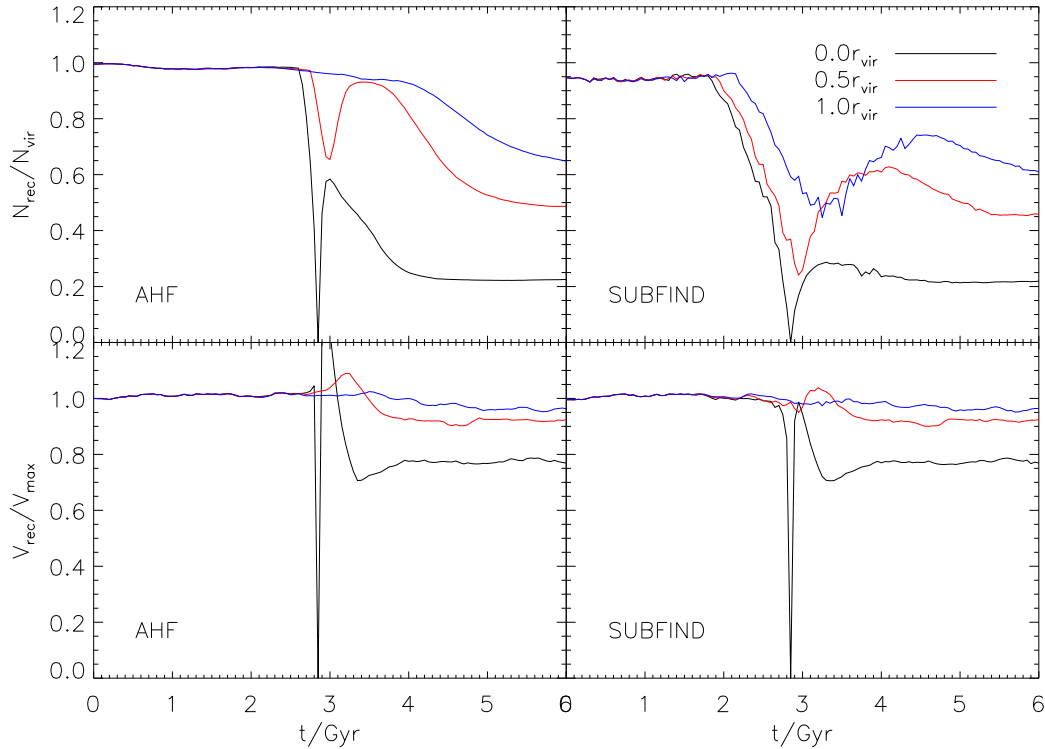}
\caption{Fraction of particles recovered (upper panels) and maximum circular velocity (lower panels) for the subhalo as a function of time as the subhalo falls through the halo.  For each case the subhalo is given a velocity along the $x$-axis toward the halo and starts offset by 3.0$r_{\rm vir}$ in the $x$-axis and 0.0 (black line), 0.5$r_{\rm vir}$ (red line) and 1.0$r_{\rm vir}$ (blue line) in the $y$-axis.  This corresponds to a closest radial approach to the centre of the halo of 0.0, 0.2$r_{\rm vir}$ and 0.5$r_{\rm vir}$ respectively.}
\label{deflection}
\end{figure*}

The second method of investigating the infall of a subhalo was to allow the system to evolve under gravity.  The same halo and subhalo properties were set up as in \S\ref{place}.  The subhalo was then placed so that $r_{\rm sep}=3r_{\rm vir}$ of the halo and it was given a velocity toward the centre of the halo from equation~(\ref{gp}).  The subhalo was then left to free-fall through the halo for 6 Gyr using \textsc{gadget}-2 with gravitational softening $\epsilon=3 \rm kpc$.  Snapshots were taken every 0.05 Gyr.  During this run cosmological expansion was turned off so the haloes were only affected by gravity.

Fig.~\ref{infall} shows the fraction of particles recovered by \textsc{subfind} and \textsc{ahf} as the subhalo passed through the halo.  The subhalo undergoes a large amount of stripping, loosing around 75 per cent of its mass.  Most of this stripping occurs as the subhalo passes through the very centre of the halo.  This corresponds to the greatest rate of change of the potential and so would be expected to have the largest effect.  As predicted in \S\ref{place} both halo finders fail to recover the subhalo as it passes through the centre of the halo and disagree about the size of the subhalo immediately either side of this region.  The largest discrepancy occurs when the subhalo is within the virial radius of the halo.  As expected due to its definition of a subhalo, \textsc{subfind} recovers a smaller subhalo during the infall phase compared with \textsc{ahf}.  After the subhalo has passed the centre of halo, \textsc{ahf} recovers a much larger number of particles due to its unbinding procedure being less efficient and this is discussed further in \S\ref{strip}.  As expected, the level of stripping observed is consistent with \citet{Hayashi03} and higher than \citet{Kazantzidis04}.

\section{Subhalo Stripping}
\label{strip}

As seen in \S\ref{dynam}, an infalling subhalo only undergoes stripping as it passes through the very centre of the halo.  This should mean that any subhalo that does not pass through the centre of the halo and is merely deflected around it should undergo significantly less stripping.  To test this hypothesis, the subhalo was placed at a separation of 3.0$r_{\rm vir}$ in the $x$-axis and 0.0, 0.5$r_{\rm vir}$ and 1.0$r_{\rm vir}$ in the $y$-axis.  In each case the subhalo was given the same velocity along the $x$-axis toward the halo as in \S\ref{dynam}.  The subhalo that was on the $x$-axis followed the same path as the subhalo in \S\ref{dynam} passing straight through the halo centre.  The other two subhaloes were deflected around the halo centre with closest approaches of 0.2$r_{\rm vir}$ and 0.5$r_{\rm vir}$ respectively.

Fig.~\ref{deflection} shows the fraction of particles recovered by each halo finder for the three scenarios outlined and also the value of the peak in the circular velocity profile.  Both halo finders give consistent values for the the final sizes of the subhalo after stripping.  For the two subhaloes that do not pass through the centre of the halo, the amount of stripping is noticeably less.  The subhalo loses around 35 per cent and 50 per cent of its mass for closest approaches of 0.5$r_{\rm vir}$ and 0.2$r_{\rm vir}$ respectively compared with over 75 per cent if it passes through the centre.

\begin{figure}
\includegraphics[width=82mm]{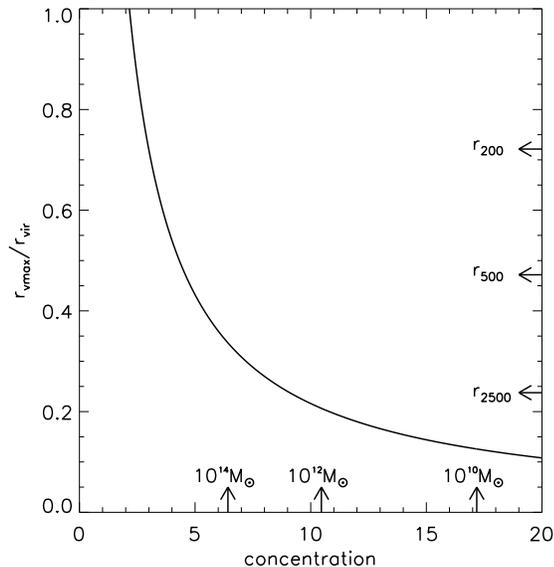}
\caption{The position of the peak of the circular velocity profile in relation to the concentration of a halo.  Typical halo concentrations from \citet{Neto07} and radial densities are also labelled.}
\label{conc}
\end{figure}

Comparing the halo finders as the subhalo passes through the central region of the halo, both show a characteristic dip in the number of particles recovered.  It is also noticeable that as the subhalo leaves the centre of the halo, \textsc{ahf} always finds a larger subhalo than \textsc{subfind}.  This is also shown very clearly in Fig.~\ref{infall} where in the region $0<r_{\rm sep}/r_{\rm vir}<1$ \textsc{ahf} gives much higher recovery of particles compared with \textsc{subfind} which has flattened off.  The cause of this difference can be seen in the lower left panel of Fig.~\ref{deflection} by considering the maximum circular velocity.  After the subhalo has passed through the centre of the halo, the maximum circular velocity recovered by \textsc{ahf} spikes meaning that background halo particles are being included in the subhalo.  There is no such spike in the \textsc{subfind} value (lower right panel).  This shows that the unbinding of particles is more efficient in \textsc{subfind} than \textsc{ahf}.  This discrepancy is caused by \textsc{ahf} assuming spherical symmetry for the unbinding when the subhalo becomes elongated in the centre of halo and is no longer a spherical shape.

For the subhalo with the closest approach of $0.5r_{\rm vir}$, \textsc{ahf} shows a smooth transition in the size of the subhalo, while \textsc{subfind} shows the size to decrease and then increase again.  During this transition the subhalo always has a finite size as the subhalo does not pass close enough to the halo centre to completely vanish.  The decrease and increase in the size of the subhalo demonstrates that it is being truncated at a radius smaller than its actual size.  As the saddle point in the density profile corresponds to the tidal radius \citep{Tormen98}, this in turn shows that a subhalo not passing through the centre of a halo will not be completely stripped down to its tidal radius.  This is perhaps not that surprising as the subhalo has not spent a long enough time in the halo to undergo the full effects of tidal stripping.

The maximum circular velocity is shown in Fig.~\ref{deflection} to be a much more stable quantity compared to particle number as expected from \S\ref{place}.  The strong radial dependence of \textsc{subfind} in particle number is not present in maximum circular velocity.  While this is an advantage in recovering properties of the subhalo, Fig.~\ref{deflection} also shows how this quantity can be misleading when considering stripping.  For the case where the subhalo passes within 0.5$r_{\rm vir}$, the subhalo was stripped of around 35 per cent of its mass, but the maximum circular velocity changes by less than 5 per cent.  This is caused by the maximum circular velocity being located at a radius much closer to the centre of the subhalo and so is less affected by stripping which occurs primarily in the outer regions.

\section{Circular Velocity}
\label{circ}

\begin{figure}
\includegraphics[width=82mm]{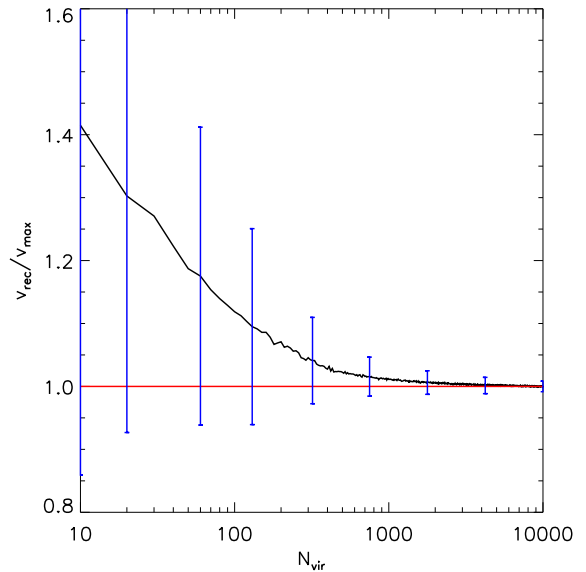}
\caption{The recovered maximum circular velocity compared with number of particles used to generate a $M_{\rm vir}=10^{12} \rm M_{\odot}$ and $c=12$ halo.  Error bars represent 1 standard deviation and are distributed symmetrically in log space.  For the average to be within 2.5 per cent of the maximum value, in excess of 500 particles are required.}
\label{res}
\end{figure}

As seen in the previous sections, the peak in the circular velocity profile of a subhalo is a more stable quantity to recover than the total subhalo mass.  The origin of this stability is related to the fact that the radius at which the maximum circular velocity is reached is located much closer to the centre of the halo and so is unaffected by truncation.  Fig.~\ref{conc} shows how the position of peak changes with the concentration of a halo.  For a NFW halo this can be obtained numerically to give,
\begin{equation}
\label{vmax}
\frac{r_{\rm vmax}}{r_{\rm vir}} \simeq \frac{2.16}{c}.
\end{equation}
\noindent The values determined by equation~(\ref{vmax}) are based on an ideal NFW halo, but for low resolution haloes there will be deviations from this curve.  For the subhalo used in this work ($c=12$) $r_{\rm vmax}=0.18r_{\rm vir}$ which corresponds to roughly $r_{5000}$ (the radius at which the enclosed density is 5,000 times the critical density, $\rho_{\rm crit}$).  Stripping occurs in the outer regions of the subhalo and so for it to affect this radius a large amount of material needs to be lost, consistent with Fig.~\ref{deflection}.

One of the main issues with using the maximum circular velocity of a halo is how its measurement depends upon resolution.  To investigate this, we generated a halo with $M_{\rm vir}=10^{12} \rm M_{\odot}$ and $c=12$ in isolation using a different number of particles within the virial radius each time.  For each number of particles within the virial radius, we constructed 1,000 realisations in order to constrain the variation.  Fig.~\ref{res} shows how the recovered maximum circular velocity varied with the total particle number.  For the sparsely populated realisations the average maximum circular velocity was higher than the analytic value.  As more particles were used, the two values converged.  For the average value to be within 2.5 per cent of the analytic value, in excess of 500 particles were required in the halo.  The variation of the maximum circular velocity between different realisations of the same total virial particle number is strong for the sparsely populated haloes.  At all points the curve is within 1 standard deviation of the analytic value, but the variation is clear where for 10 particles the standard deviation is 0.56 compared with 0.002 for 10,000.  To obtain an accurate value for the maximum circular velocity of a recovered subhalo, its resolution has to be taken into account.

\section{Summary and Conclusions}

Halo finders are an important tool for the analysis of cosmological simulations.  They are pivotal in the construction of merger trees, which underpin galaxy formation modelling, and their results allow us to characterise, for example, the abundance and spatial distribution of both dark matter haloes and subhaloes.  There are as many techniques for identifying haloes and subhaloes in cosmological simulations as there are halo finders and so it is interesting to ask whether or not (sub)halo properties recovered by different halo finders are consistent.

In this paper we have compared and contrasted the results of two halo finders, \textsc{subfind} and \textsc{ahf}, that use fundamentally different approaches to identifying subhaloes.  We have taken a simple test problem, the identification of a NFW subhalo embedded in a more massive NFW halo, and compared the performance of \textsc{subfind} and \textsc{ahf} in recovering the mass of the subhalo at different radii within its host.  As shown using \textsc{subfind}, halo finders that identify subhaloes as overdensities will have a strong dependence on the local density.  This is demonstrated in the strong radial dependence in the fraction of a model subhalo \textsc{subfind} recovers.  As the subhalo gets closer to the centre of the halo, the background density from the halo is rising.  With a higher background density and the same density for the subhalo, the overdensity will be less leading to a smaller subhalo being recovered.  By the time the subhalo is in the centre of the halo, which corresponds to the densest point, the overdensity becomes negligible leading to no saddle point and the subhalo is no longer detected.  While the size of the overdensity recovered roughly corresponds to the tidal radius of the subhalo, it has been shown that not all subhaloes are stripped down to this size when they pass through a halo.  The authors of \textsc{subfind} are aware of these issues \citep[see \S4.1 of][]{Springel08} and post-process, but where this effect is not taken into account it could have profound consequences on substructure studies.

The radial dependence of locating subhaloes as overdensities will have a large effect on measures of tidal stripping.  As a subhalo plunges into a halo, the halo finder will reduce the size of the subhalo due to the increase in density.  If this is not considered, then it will appear the subhalo is undergoing a larger amount of stripping as it falls through the halo than it actually underwent.  Stripping will be further complicated by the fact it occurs in the outer region of the subhalo, an area that is not included in the truncated subhalo that is recovered.  This can lead to confusion when comparing the recovery of \textsc{ahf} and \textsc{subfind}.  \textsc{ahf} indicates that most of the stripping occurs as the subhalo passes through the centre of the halo and not during the infall, but \textsc{ahf} has been shown to have inefficient unbinding causing it to retain a larger fraction of particles.  Meanwhile \textsc{subfind} indicates a more gradual process, but the effects of truncation will cause the recovered subhaloes to always be lower estimates of the size.  Further studies will need to be made to determine how dramatic the effect of stripping is on an infalling subhalo.

The radial dependence in recovery will also have important implications for the subhalo mass distribution.  Two subhaloes that have identical mass can be recovered with different sizes based on position.  This will lead to large subhaloes being recovered as smaller ones, in turn, leading to subhalo mass distributions biased towards the low mass end.  Whilst most subhaloes that reside in the inner region of the halo will have undergone a large amount of stripping and will be smaller anyway, the effect of truncation still needs to be considered alongside the underlying physics.  These issues highlight that the recovered mass identified using the overdensity method is not a good property to consider when studying subhaloes.  This is true even as far out as the virial radius of the halo, where the mass can be underestimated by around 25 per cent.

A more stable quantity to consider is the peak in the circular velocity profile.  This is located much closer to the centre of the subhalo and so will be less affected by truncation and the particular choice of the definition for an entire subhalo.  Both \textsc{ahf} and \textsc{subfind} recover consistent values for the maximum circular velocity at all radii within the halo, except at the very centre of the halo where no particles are recovered.  This makes the circular velocity peak a useful quantity to track subhaloes and gives a good indication of initial mass.  However, when considering stripping, the circular velocity peak is no longer useful.  Being located so close to the centre of the subhalo, a substantial amount of the outer layers can be stripped before the peak in the circular velocity is affected.

Two methods of improving the accuracy of subhalo recovery would be halo tracking and phase space.  Halo tracking involves identifying the subhalo before it falls into the halo so all the particles that were originally part of the structure are followed and at each time step they can be tested to see if they are still part of the substructure.  The disadvantage of this technique is that it requires multiple snapshots to identify the subhalo, not a problem for the second method of phase space.  Phase space takes into account not only the spacial position of the subhalo particles, but also links particles based on a common velocity.  By considering haloes in phase space density, any subhaloes that are present will stand out as overdensities.  These can then be isolated.  For subhaloes in the centre of the halo, the difference in the bulk velocity of the particles would cause them to be separated in phase space.  The only remaining problem would be if a subhalo was at rest in the centre of the halo.  These structures could not be separated in phase space, but it is arguable whether such a structure would be a dynamically independent entity.

\section*{Acknowledgments}

The authors wish to thank Alexander Knebe, Steffen Knollmann, Justin Read and Volker Springel for useful discussions.  SIM and FRP would also like to thank the \textsc{astrosim} network of the European Science Foundation (Science Meeting 2910) for financial support of the workshop 'Haloes going MAD' held in Miraflores de la Sierra near Madrid in May 2010.  CP acknowledges the support of the \textsc{stfc} theoretical astrophysics rolling grant at the University of Leicester.  This research made use of the High Performance Computing (HPC) facilities at the University of Nottingham.

\bsp

\label{lastpage}

\end{document}